# Superlattice Structures in Twisted Bilayers of Folded Graphene

Hennrik Schmidt[‡ 1,2], Johannes C. Rode[‡ 1], Dmitri Smirnov [1], and Rolf J. Haug [1]

[1]Institut für Festkörperphysik, Leibniz Universität Hannover, 30167 Hannover, Germany

[2]Graphene Research Centre, National University of Singapore, 117546 Singapore, Singapore

[‡]These authors contributed equally

## Abstract

The electronic properties of bilayer graphene strongly depend on relative orientation of the two atomic lattices. While Bernal stacked graphene is most commonly studied, a rotational mismatch between layers opens up a whole new field of rich physics, especially at small interlayer twist. We here report on magnetotransport measurements on twisted graphene bilayers, prepared by folding of single layers. These reveal a strong dependence on the twist angle which can be estimated by means of sample geometry. At small rotation, superlattices with a wavelength in the order of 10 nm arise and are observed by friction atomic force microscopy. Magnetotransport measurements in this small angle regime show the formation of satellite Landau fans. These are attributed to additional Dirac singularities in the band structure and discussed with respect to the wide range of interlayer coupling models.

## Introduction

Two-dimensional monolayer graphene (MLG) exhibits outstanding intrinsic electronic properties[1], paired with considerable possibilities of manipulation due to high robustness and accessibility at the surface. Furthermore, its atomically thin nature makes it extremely sensitive to ordered as well as disordered potential fluctuations in its vicinity. To this end, several groups have recently observed

Hofstadter´s butterfly[2] in the spatially modulated potential landscape of graphene on hexagonal boron nitride (hBN)[3-5]. Equally fascinating effects can be expected of twist induced moiré superstructures (Fig. 1a) in rotationally mismatched graphene double-layers[6-9]. Apart from the common Bernal-stacked variety[10], these twisted bilayers (TBG) constitute a whole new field of their own: Layers of large rotational mismatch effectively decouple[11-13], exhibiting reduced Fermi velocities for decreasing interlayer twist in many cases[11,14-17]. At the smallest angles, totally different electronic structures are expected[7,9,15,18]. In recent years, TBG of various angles have been grown[19] and optical studies[20,21] as well as scanning tunneling spectroscopy[11,22-24] were performed on samples of different interlayer twist, revealing e.g. low energy van Hove singularities and charge density waves. However there is few systematic work on electronic transport so far, focusing only on large[17,25,26] or disordered small angle systems[27]. Here we present a study on high quality folded graphene monolayers of different twist angles $\theta$, the smallest of which lead to novel transport features in the form of satellite Landau fans, caused by twist induced long-wavelength superlattices.

## Results

**Folded graphene samples.** Our folded layers are obtained by mechanical manipulation via Atomic Force Microscope (AFM) or incidental flip-over during the exfoliation of natural graphite. While the method of micromechanical cleavage is known to yield flakes of high crystalline order, a successive folding step induces little further contamination between layers, thus providing TBG of the highest quality. A further advantage of folded samples is that from sample geometry alone, the twist angle $\theta$ can be estimated (see Fig. 1c): As graphene is most commonly terminated by armchair- or zigzag-edges, alternating in 30 degree steps[1,28], an according set of straight edges provide a crystallographic reference direction. Interlayer twist then relates to the angle $\varphi$ between this reference and the folding edge by $\theta = 2 \cdot \varphi$. Note that this principle is not applicable without the common folding edge between layers. Due to graphene´s sixfold symmetry, $\theta$ may be projected into the range of $0° < \theta \leq 30°$ (see methods section).

In case of the shown optical image (Fig. 1b), the twist angle can thus be narrowed down to 1.5 ° +/ 0.5. Figure 1d shows AFM topography data of the same sample. Mono- and bilayer regions can clearly be distinguished, as analyzed in the cross-sections in Fig. 1e. It is worth noting, that we could not observe the formation of bubbles in the folded areas as can be found in samples fabricated by transfer methods[29]. This points toward little contamination between clean interfaces. Note also that at the folded edge of the TBG a small but distinct elevation of 2.5 Å is present (grey circle, Fig. 1e), indicating a bended but unbroken interconnection between layers which may contribute to electronic transport, as discussed later.

The two rotated lattices can arrange in periodic superstructures reproducing the original honeycomb pattern on a twist dependent length scale of

$\lambda(\theta) = a / [2 \sin(\theta / 2)]$ (1),

$a$ being the length of graphene´s lattice vector [6,8,15]. The alternating interlayer registry (sublattice AB, BA and AA) modulates coupling and potential landscape in the TGB which enables experimental visualization of moiré structures in graphene by scanning probe microscopy[22,24,30]. As recently shown[30], friction AFM serves as a convenient tool for resolving large period superlattices in van der Waals heterostructures. Figure 1f shows a lateral force microscopy scan of the TBG area indicated by the black box in Fig. 1d. The dashed white star marks three distinct symmetry directions in the friction force plot. These are clearly confirmed by the prominent hexagonal pattern in the Fourier transform (Fig. 1h) which points to a trigonal lattice of period $\lambda \sim 9$ nm. Figure 1g shows a close up of the yellow box in panel f with accordingly added unit cells and lattice vector $\lambda$. Using eq. 1, the resolved lattice matches the moiré structure to an interlayer twist of $\theta \sim 1.6$ ° which fits the geometrically estimated value of 1.5 ° +/- 0.5 very well. Note that the large scale corrugation in Fig. 1g stems from inherent roughness of the underlying $SiO_2$ substrate which is amorphous and unlike hBN does not lead to any periodic superstructures.

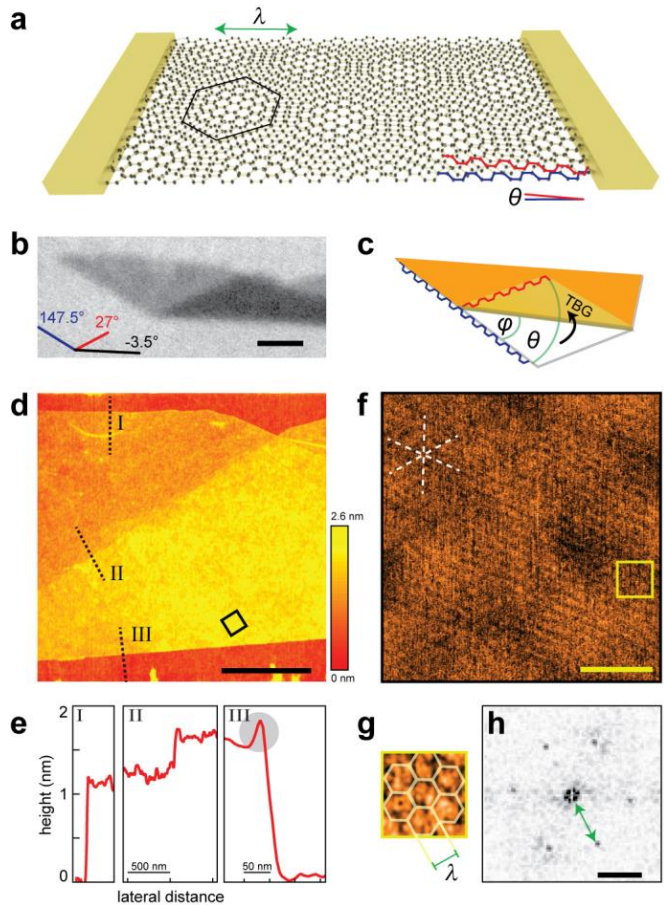

**Figure 1 | Twisted bilayer graphene in folded samples.** (a) Sketch of a contacted twisted bilayer graphene sample with a rotational mismatch $\theta$ between lattices, indicated by the red and blue armchair edges of the two layers. A hexagonal moiré pattern with angle dependent wavelength $\lambda$ emerges. (b) Optical image of a folded single layer of graphene. Colored lines indicate the orientation of the different edges. The scale bar is 2 μm. (c) From the orientation of the original edges and the folding edge, the rotational mismatch can be estimated. Crystallographic edges (interchangeable) and the folding angle $\varphi$ as well as the resulting twist angle $\theta$ are highlighted. Note that $\theta$ may be projected to $\theta \in (0°, 30°]$. (d) AFM topography map of the folded sample shown in the optical picture. The dotted lines indicate the position and direction of the cross-sections depicted in panel (e), the black box approximately marks the area of the friction AFM scan in (f). Red color corresponds to low, yellow to high topographic features, as indicated by the color scale bar. The black lateral scale bar indicates 1 μm. (e) Averaged cross-sections over a small area AFM topography scan at intersections of substrate to monolayer graphene (I), monolayer to twisted bilayer (II), and twisted bilayer to substrate (III). In (III) an additional elevation due to the folding can be seen close to the edge (grey circle). (f) Friction AFM image of the TBG. Bottom right corner of the panel corresponds to the bottom corner of the black square in panel (d). The dashed white star indicates symmetry directions of the superlattice, the yellow box approximately marks the area of a close up in panel (g). The scalebar is 50 nm. (g) Close up of the yellow square in panel (f) with overlain moiré unit cells. (h) Fourier transform of the lattice in panel (f) showing hexagonally arranged peaks corresponding to a trigonal lattice with a 9 nm wavelength. The scalebar is 100 μm$^{-1}$.

**Magnetotransport for different rotational mismatch.** To investigate the electronic properties of graphene bilayers with different stacking, magnetotransport measurements in perpendicular fields up to $B = 13$ T were performed. Fig. 2 shows examples for AB-stacking, small twist angle and larger angle. For each case, the resistance as function of gate voltage and magnetic field (middle row) as well as a cross-section at fixed charge carrier density over $B^{-1}$ (top) and the field effect curves at $B = 0$ T (bottom) are shown. First we focus on the limits of zero and large relative twist angle: In case of Bernal-AB-stacking (Fig. 2, left), Landau level spectrum and phase of the Shubnikov-de Haas oscillations in $B^{-1}$ show a Berry's phase of $2\pi$ [10]. For larger rotational mismatch ($\theta > 3°$), the layers effectively decouple and linear monolayer spectra are recovered[12-14], while interaction may be seen in a reduction of Fermi velocity which is strongly angle dependent[11,14,15,17]. For the example in the right column of Fig. 2, the Fermi velocity is found to be 75 % +/- 5 of the original value $v_F = 10^6$ m s$^{-1}$ [31], as obtained from temperature dependent Shubnikov-de Haas measurements. These reduced values are used to estimate the rotational mismatch as $\theta = 3.25° $ +/- 0.75 according to theoretical considerations[15]. In between those cases at approximately $0.3° < \theta < 3°$ [9] lies a small range of twist angles, where rich physics like the Hofstadter butterfly are expected to be observable[6-9]. Indeed our transport data are most complex for this small angle regime: In the depicted example (Fig. 2e), a main Landau fan is found to be originating from around - 15 V backgate voltage. The Shubnikov-de Haas oscillations in panel b indicate charge carrier transport through a graphene monolayer (Berry's phase of $\pi$). In addition to this, a second fan arises at a backgate voltage of around 30 V. At zero magnetic field, it is accompanied by a small but notable shoulder in resistance and a dip in the otherwise linear conductance (Fig. 2h). This can be attributed to a large wavelength moiré pattern between the two rotated hexagonal lattices (Fig. 1a, f-h) which induces a periodic potential landscape on a twist dependent length scale. Bragg scattering by the correspondingly small superlattice Brillouin zone will lead to replica satellite Landau fans at higher energies[3-5], while periodically alternating interlayer coupling should result in a more complex electronic spectrum[6,7,9,32-34].

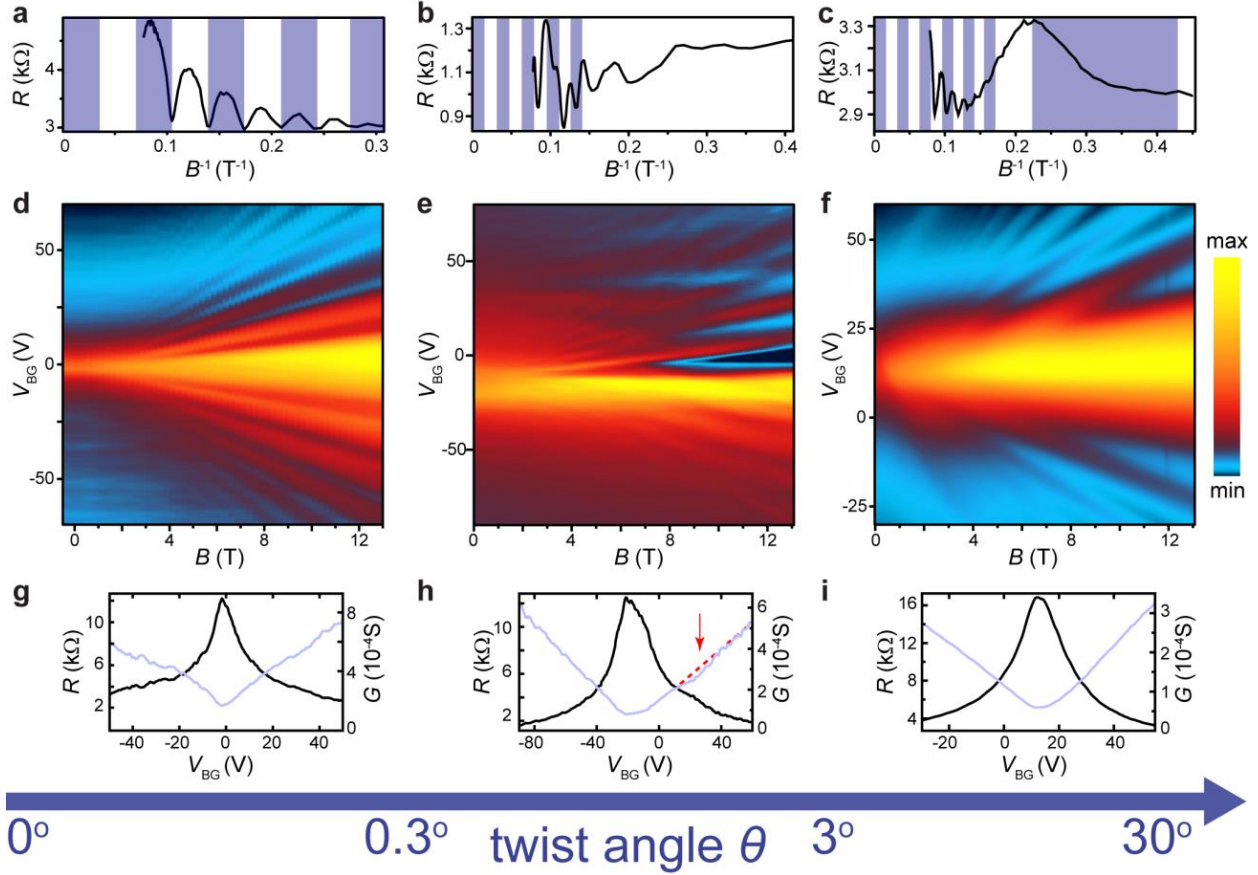

**Figure 2 | Transport characteristics in bilayer graphene of different angles.** Resistance as a function of gate voltage and magnetic field for different bilayer devices with **(d)** Bernal stacking, **(e)** a small rotational mismatch $\theta = 1.5°$ +/- 0.5 and **(f)** a mismatch of $\theta = 3.25°$ +/- 0.75. The plots above show the according characteristic Shubnikov-de Haas oscillations for the three cases at constant backgate voltages of $V_{bg}$ = + 40 V, + 89 V and + 60 V respectively. The blue-striped backdrops highlight **(a)** the typical Bernal-bilayer sequence with Berry's phase of $2\pi$ (color changes at every minimum), **(b)** the monolayer sequence with Berry's phase of $\pi$ (color changes at every maximum) of the main Landau fan and **(c)** two independent monolayer sequences for different charge carrier densities. The bottom row shows resistance and conductance vs. backgate voltage at zero magnetic field. For the small angle sample in panel **(h)** the field effect shows a dip below the expected linear conductance trend (red dotted line) and a shoulder in resistance in the region of the additional Landau fan in panel **(e)**. The blue arrow at the bottom marks the angular range for the predicted different coupling regimes.

**Small twist angles.** In the following we focus on four TBG samples in the range of $0.3° < \theta < 3°$. Figures 3a and b show the differential resistance $\delta R / \delta B$ for two such small angle devices. Originating from the charge neutrality point (CNP), a monolayer Landau fan with minima at filling factors $\nu = n / [B / (\Phi_0)] = 2 + 4i$ is present ($n$ is the charge carrier density, $\Phi_0 = h / e$ the magnetic flux quantum and $i$ an integer). Additional satellite fans can be identified at $n \sim -1.5 \times 10^{12}$ cm$^{-2}$ (Fig. 3a) and $n \sim -3.5 \times 10^{12}$ cm$^{-2}$ (Fig. 3b) respectively. Especially in panel b, somewhat less defined features arise in between at intermediate charge carrier concentrations. This could either hint to a more complex coupling or stem from limited long range order, leading to deviations in the moiré period.

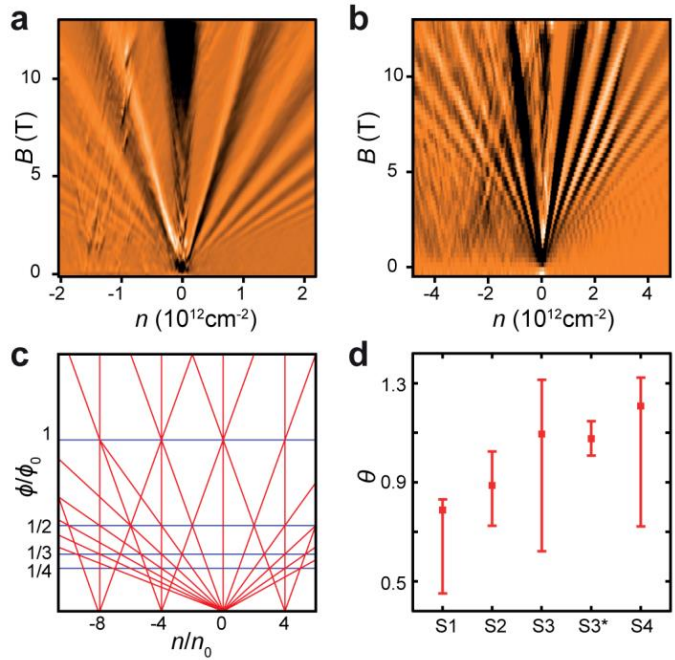

**Figure 3 | Relation between satellite fans and unit cell area in TBG. (a)** and **(b)** Differential resistance as a function of magnetic field and backgate voltage of two samples (*S1, S4*) with different small rotational mismatch, both showing the main Landau fan at around $n = 0$ and an additional satellite fan in the regime of hole conduction. **(c)** Schematic of the expected Landau-fan sequence in the model of periodic superlattice potential fluctuations. The *x*-axis scales in charge carrier density $n$ normalized by $n_0 = 1 / A$, the *y*-axis in magnetic flux through superlattice unit cell $B \cdot A$ normalized by the magnetic flux quantum $\Phi_0$. Neighbouring fans intersect at reciprocal multiple integers of $\Phi_0$. **(d)** Twist angles $\theta$ as obtained from the satellite-fan distance for different samples. Error bars originate from uncertainty in inter-fan distance and capacitive coupling. *S3* and *S3\** indicate the values for the same sample before and after annealing.

Following the simple model of an additional superlattice potential, the shift in *n* is dependent on $\theta$ and given by

$$n \cdot A = 4 \quad (2),$$

where the moiré unit cell (area *A*) of the fourfold degenerate system gets filled [4]. The expected Landau level picture is mapped out in Fig. 3c. From the transport data of different small angle samples, twist angles are determined by means of equations 1 and 2 and the relation

$$\lambda = \sqrt{2A / \sqrt{3}} \quad (3)$$

between wavelength and unit cell area of the hexagonal moiré superstructure. Thusly extracted twist angles are summarized in Fig. 3d for samples *S1* to *S4*: The two values for sample *S3* have been extracted from transport measurements before and after (*S3\**) annealing, which changed the overall doping level and notably increased the clarity of the secondary fan but did not affect inter-fan distance.

A more detailed analysis of sample *S3\** is given in Fig. 4. Panel a shows the measured magnetoresistance vs. charge carrier concentration and magnetic field. Two Landau fans with origin at charge neutrality and about $n = 2.75 \times 10^{12}$ cm$^{-2}$ respectively can be observed and are fit by a set of slopes similar to the schematic in Fig. 3c. The good match confirms a systematic nature of the observed features. Interestingly in this sample, only every second slope of the main fan seems to be found in the satellite one as becomes evident at the intersections with the black horizontal lines. While the main fan is described by fourfold degeneracy as extracted from the expected capacitive coupling constant, the Landau levels of the satellite one are thusly eightfold degenerate. This can be verified on panel b, which depicts our resistance data against backgate voltage and inverse magnetic field: Whereas the main fan crosses Landau levels of the second one at every equally spaced horizontal line, the satellite fan only intersects maxima of the first fan at either dashed or solid lines. This coexistence of different degeneracies in the presented example exceeds the simple model of a superlattice potential applicable to graphene on hBN[3-5] and might hint towards a more complex coupling in TBG. Figure 1c shows the differential resistance vs. charge carrier density and magnetic field revealing traces of a further Landau fan at higher energies. Note that the origins of the observed Landau level sequences are roughly equally spaced on the axis of carrier density which points to a systematic relation of the third one to the ones examined above.

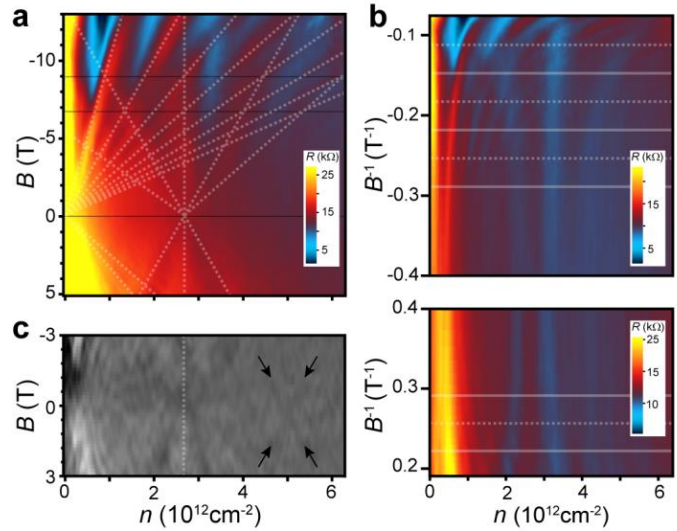

**Figure 4 | Detailed analysis of the satellite fan feature.** (a) Resistance of a small-angle folded sample (*S3\**) as function of magnetic field and backgate voltage. The dotted lines trace main (left) and satellite fan (right), the black horizontal lines indicate Landau-fan intersections. (b) Resistance over the inverse positive and negative magnetic field of polarity. The additional features can be seen for both polarities. Horizontal lines mark intersections: The satellite fan intersects Landau levels of the main one at every second line (either dashed or solid only), indicating doubled degeneracy. (c) In the derivative of the resistance, a third fan can be identified as indicated by the arrows. The dotted vertical line marks the position of the second one.

**Transport over the folded edge.** While coupling in the plane leads to the demonstrated novel electronic properties, also the edges play an important role, especially in case of folded samples featuring a bended interconnection as depicted in the inset of Fig. 5b. Figure 5 shows transport data of a device, in which two monolayer regions were separately contacted on both sides of the folding, thus forcing the charge carriers to either hop between layers or pass the curved folding edge. Such an edge (as observed in the AFM cross-sections in Fig. 1e) induces strong gauge fields[35] and also will exhibit different charge carrier densities. Additionally to the usual CNP peak (filled circle) and features related to a satellite fan at positive gate voltages (white lines) another peak can be observed at around - 25 V. This maximum is independent of the magnetic field $B$ applied perpendicular to the sample plane. As the average bending area will be parallel to $B$, it is tempting to attribute these feature to the folded edge and the predicted formation of snake states[35].

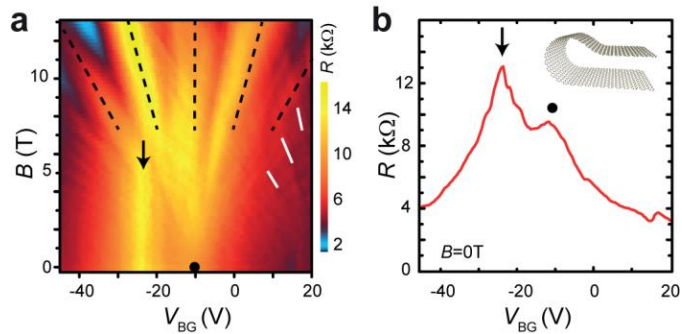

**Figure 5 | Transport contributions of the folded edge**. **(a)** Resistance of a folded sample in the vicinity of the main Landau fan (indicated by the black lines) measured over the folding edge. While at positive gate voltages features of the satellite fan can be seen (white lines), an additional peak at $V_{bg}$ = - 25 V can be identified, which is independent of magnetic field. **(b)** Resistance at $B$ = 0 T. The charge neutrality point of the TBG sample (circle) and the additional edge-related peak (arrow) can be distinguished.

## Discussion

We have studied folded double layer graphene devices of various twist angles. At small rotational mismatch, the formation of superlattice structures is observed by AFM while systematic satellite Landau fans emerge in transport measurements. These indicate the generation of secondary Dirac singularities in the band structure at experimentally accessible energies comparable to observations on heterostructures of closely aligned graphene on hBN[3-5]. While underlining the important role of interlayer twist, our observations also point to the requirement of long range order and a well defined moiré period over a large sample area, as discussed later on. A first intuitive measure for the visibility of superlattice effects in transport measurements is the mean free path $l_m$ in relation to the moiré period $\lambda$, which is proportional to the charge carrier mobility $\mu$, readily derived from the field effect measurements at zero magnetic field. The examined samples exhibit typical values of $\mu \sim 2500$ V s cm$^{-2}$, corresponding to a mean free path of $l_m = 46$ nm at a charge carrier concentration of $2.5 \times 10^{12}$ cm$^{-2}$. As $l_m$ exceeds the superlattice period which is around 10 nm in the examined samples, the requirements for an observation in transport should be given. The superlattice effects in graphene on hBN[3-5] are significantly more pronounced than in our systems which can be explained by the high charge carrier mobility in devices on a smooth hBN substrate, exceeding the one in our systems by at least an order of magnitude[4]. While the mean free path may determine visibility and sharpness of the superlattice features, another factor seems to be responsible for clarity and definition of the observed satellite Landau fans. This becomes evident for sample *S3* which has been measured before (*S3*) and after (*S3\**) an additional outside annealing step. The satellite features before had a frayed appearance, similar to the examples in Fig. 3a and b. Afterwards, a well defined Landau fan emerges as depicted in Fig. 4, while the mobility remains virtually unchanged. A possible explanation for this would be an improvement in global interlayer registry: As local strain can result in slight variations of interlayer twist over the sample area, causing comparatively large deviations in moiré unit cell (eq. 1) and electronic structure, the annealing process and subsequent cooling may have led to a more homogenous interlayer registry thus improving the clarity of superlattice transport signatures. In a similar vein, as such signatures have not been reported on epitaxial grown bilayers (which often exhibit small twist angles around 2 ° [12,27]), the high crystal quality of exfoliated and folded samples is likely to play a crucial role in the observed superlattice physics.

Although the observed transport signatures as well as a large wavelength moiré superstructure between two hexagonal lattices constitute striking similarities to the hBN case, the physics in a graphene bilayer are somewhat more complex: In addition to the common superlattice potential, there exists a periodically alternating coupling between the two electrically active graphene layers. In the absence of coupling, the description of a TBG in momentum space is given by two rotated sets of Dirac cones displacing the $K$-points of two layers by $\Delta K = 2\,K\sin(\theta/2)$ ($K$ being the magnitude of the vector **K** to the Dirac point)[15,32]. Accounting for interlayer coupling, theoretical models predict an angle dependent reduction of Fermi velocity $v_F$ and merging of the displaced Dirac cones in low energy van Hove singularities (VHS) [14-16]. Landau quantization in such a system would yield a single layer sequence of doubled degeneracy below and a Bernal bilayer sequence above the VHS [36, 37]. The energy $E_{VHS}$ at the singularity and the transition in quantization can be estimated as $E_{VHS} = v_F\,\hbar\,\Delta K / 2 - t_\theta$, $v_F$ being the Fermi velocity and $t_\theta$ the interlayer hopping amplidude [14,22,24]. While the latter may vary greatly, depending on the system [22], a value of $t_\theta \cong 0.1$ eV seems to be a good estimation[14,24]. Using the single layer relation between charge carrier concentration $n$ and energy $E$, $n = E^2 / (v_F^2\,\hbar^2\,\pi)$, the singularity should occur around $n = 2\,E_{VHS}^2 / (v_F^2\,\hbar^2\,\pi)$ (the additional factor 2 accounts for the extra twofold degeneracy below the VHS). For the device exhibiting a larger twist angle, as presented in the right column of figure 2, the experimentally probed range lies well below this transition which would be around $V_{bg} = 107$ V above or below charge neutrality. This is in agreement with the observation of single layer behavior in the measurements, while stronger influence of the backgate in the lower sheet leads to asymmetric charge distributions and separation of two distinguishable Landau sequences. Together with an observable reduction in Fermi velocity, these measurements are in accordance with the above model. However, looking at the systematic formation of multiple Landau fans in the small angle regime, another explanation is needed. Chu et al. [32] point out the possibility of the coexistence of interlayer coupling phenomena and superlattice Dirac points generated by periodic potential fluctuations like in graphene on hBN. Mele et al. describe a coupling mechanism which suppresses the renormalization of Fermi velocity as well as the formation of a VHS, producing second generation Dirac singularities at the crossing points instead[33,34]. The latter model is of special interest for the case of very small angles, since the scenario of a continuous reduction in $v_F$ is no longer valid below a certain angle (depending on the model[15,16]) leading for example to flat bands and a localization of charge carriers[15]. Both models[32,33] could explain the interesting observation of multiple Landau fans and the coexistence of different degeneracies in one spectrum as suggested in Fig. 4. Our findings bear further evidence to the rich and complex physics in twisted graphene bilayers especially at the smallest angles, encouraging more theoretical and experimental work on the topic. Moreover the presented transport signatures suggest that an emergence of electronic superstructure effects in double layer systems is not limited to ultraclean graphene-boron nitride samples and could also be expected to appear in other upcoming van der Waals heterostructures[38,39].

# Methods

**Sample preparation.** Our samples were prepared by exfoliation of natural graphite and placed onto Si / SiO$_2$ substrate. Monolayers and flipped over twisted bilayers were selected via optical microscopy. A Multimode AFM with diamond coated tip of high spring constant (about 40 N m$^{-1}$) was utilized to fold further samples by programmed tip movement over the sample edges.

Devices where processed using standard electron beam lithography techniques to define and evaporate chromium-gold contacts.

Depending on geometry of the folded sample, contacts were placed in varying positions to the folded edge, which led to different contributions of the bend (the strongest being discussed in Fig. 5).

Outside annealing under argon atmosphere at 400 degrees, prior to the measurements as well as in situ annealing under vacuum inside the cryostat has been carried out.

**Atomic Force Microscopy.** The resolution of the moiré superlattice as shown in Figure 1f was acquired by friction AFM. A Multimode II with a J-type scanner was used in ambient conditions at a constant, regulated temperature. Several hours of pre-scanning were performed to reduce thermal drift. Best results were obtained with triangular Pyrex-Nitride probes of spring constant $k \sim 0.3$ N m$^{-1}$ and tip radius $r \leq 10$ nm. A low setpoint in the attractive regime was chosen at zero proportional gain and close to zero integral gain. Piezo xy-offset was held at zero. A scan rate of 10 Hz proved to be high enough to reduce thermal noise and low enough to keep piezo xy-oscillations small. The depicted example in Fig. 1f shows the lateral deflection signal in retrace direction at a scanning angle of 50 °.

**Measurement setup**. The transport measurements were performed in a $^4$He bath cryostat using DC and low frequency AC-setup. A source current was driven through the sample and voltage measured in two- and four terminal setups. The measurement shown in Fig. 5 was acquired at 2 K, all other shown data were acquired at 1.5 K. Measurements on the Bernal bilayer presented in the left column of Fig. 2 as well as the small angle sample *S4* presented in Fig. 3b were conducted in two terminal configuration. All other measurements were performed in a multi-terminal setup. While for sample *S1*, *S2*, *S4* at least one of the contacts lies on the double layer system, the device geometry of *S3* is in such a way, that two contacts lie on each side of the folded monolayer, allowing four terminal measurements across the folded edge.

**Analysis.** The capacitive coupling constant $\alpha$ between sample and backgate is estimated via the parallel-plate capacitor model. Thickness ($d = 330$ nm) and dielectric constant ($\varepsilon_r = 3.9$) of the oxidized silicon top layer on the used substrates yield a value of $\alpha = 6.53 \times 10^{10}$ V$^{-1}$ cm$^{-2}$. Fitting the observed Landau level sequences to $i \cdot g = \alpha \, \Delta U / [B / (\Phi_0)]$ with $g$ as factor of degeneracy, $g$ is readily determined, as applied to Fig. 4a.

Due to the sixfold symmetry of graphene, a twist angle $\theta'' \in [0\,°, 360\,°)$ can always be projected into an interval of $\theta' \in [0\,°, 60\,°]$ by $\theta' = \theta''$ mod $60\,°$. While, depending on the axis of rotation, a $60\,°$ twist can mean the difference between AB- and AA-stacking, there is small difference in the overall moiré pattern between $\theta'$ and $60\,° - \theta'$ for $0\,° \neq \theta \neq 60\,°$ in the low angle regimen[15], allowing for a further projection to $\theta \in (0\,°, 30\,°]$ by $\theta = 60\,° - \theta'$ for $\theta' > 30\,°$.

ACKNOWLEDGMENT

This work was supported by the DFG within the Priority Programme "Graphene"
and by the NTH School for Contacts in Nanosystems.
J. C. R. acknowledges support from Hannover School for Nanotechnology

AUTHOR INFORMATION

These authors contributed equally to the work: H. S. and J. R.

R. J. H., H. S., and J. C. R. designed the experiment.

H. S., D. S. and J. C. R. prepared the samples and carried out the measurements.

All authors analyzed and discussed the obtained results.

All authors wrote the manuscript.

Corresponding author: Hennrik Schmidt, E-mail: h.schmidt@nano.uni-hannover.de